# Computer simulation approach to reliability and accuracy in EXAFS structural determinations

Paolo Ghigna, Melissa Di Muri, and Giorgio Spinolo

Dipartimento di Chimica Fisica, INCM and C.S.T.E./CNR, Università di Pavia

I 27100 - Pavia (Italy)

## Synopsis

The fit of a set of simulated noisy EXAFS spectra samples the frequency distribution of structural parameters and gives their statistical estimators (mean, dispersion, correlations).

## Abstract

The frequency distribution of different parameters of an EXAFS spectrum can be directly sampled by analysing a population of simulated spectra produced by adding computer-generated noise to a reference pattern.

The procedure gives statistical estimators of the parameters obtained with different data processing strategies to test the performance of a strategy, to evaluate the bias introduced by random noise, and to clarify the amount of information actually contained in an experimental spectrum.

Results are given for the two simple local structures of an Ag atom surrounded by two oxygens or by six iodines.

# 1. Introduction

EXAFS (Extended X-Ray Absorption Fine Structure) is well recognised as a powerful and now popular tool for investigating local structural features in different kinds of materials. Strong points of the technique are its well assessed theoretical background [see, for instance, Rehr 2000], the ability to provide information on the environment of a chemically defined scattering centre, and the possibility to deal with different kinds of materials, from liquids or disordered solids to well crystallised samples. The popularity of EXAFS determinations has also greatly increased in recent years because of the rapid diffusion of specialised software [Filipponi DiCicco & Natoli 1995, Filipponi & Di Cicco 1995, Binsted 1998, Rehr 1994 and recent versions of the FEFF programme].

Aim of this paper is to give a first contribution to the problems concerning accuracy and reliability of these interesting determinations. Indeed, these aspects have received so far only little attention and, to the knowledge of the present Authors, only a few papers [Incoccia and Mobilio, 1984, Lytle *et al.*, 1989, Filipponi, 1995, Curis and Benazeth, 2000, Krappe and Rossner, 2000,] are available on this regard in the literature. Indeed, there are many particular aspects which make not so straightforward a conventional error analysis of EXAFS data processing. To list just the most obvious ones, there is first the effect of preliminary processing steps such as pre-edge and post-edge background removal, which heavily rely on empirical models, and on some amount of subjective evaluation. Another source of error is possibly related to the frequent practice of combining *k*-space fit with Fourier-space analysis and windowing, and also to the need of introducing some threshold value above which the coordination shells are not taken into account. Finally, different *k*-space weighting schemes are typically applied not only for the purpose of data presentation and inspection, but are also used in data processing, a practice which does not seem based on sound principles. Other difficulties, which are indeed typical of many full-spectrum structure-based data fit procedures, are due to the strong non-linear relation between experimental data and model parameters, and to the high degree of correlation between model parameters. This can reasonably produce bias and non-parabolic errors.

The approach that is used in this paper has been applied two years ago by the same group to the problem of accuracy and reliability of powder pattern structure refinement [Dapiaggi *et al.* 1998]. Only the most general ideas and the specific details will be discussed here, while reference is made to the previous work for a deeper discussion of the method.

In general, aim of the approach is to know what amount of information is actually contained in an EXAFS spectrum and can be easily retrieved by routine work and, on the contrary, what information is deeply masked by noise (and by the particular structural model) and therefore requires a specialised strategy in data acquisition and analysis, or cannot be retrieved by whatever kind of data treatment.

This task is achieved, as suggested in the well known book by Press *et al.* (1988), with computer simulated experiments. According to this approach, a ‘large’ number of synthetic data sets is prepared: each set of data differs from the others because of random errors added by computer simulation (according to assumed statistical distribution laws) to the same reference spectrum built from an assumed structural model and assumed (‘true’) parameters. Fitting all data sets to the model gives a population of parameters, and the frequency distribution over the computer generated population is used as a numerical sampling of the underlying probability distribution.

# 2. Outline

As a general rule, we kept things as simple as possible. Consequently, some problems of a real EXAFS local structure determination have not been investigated in this preliminary approach. To quote just two well-known problems, we have considered only very simple models containing a single coordination shell, and we kept simple both the pre-edge and post-edge backgrounds. Broadly speaking, the present paper investigates only the effect of random noise.

## 2.1. Reference XAS spectra

The reference spectra have been deterministically produced from assumed models and from a given set of ‘true’ parameters. Two different models are discussed in the present work. In the first one, the photoabsorber (Ag) is surrounded by a small number (two) of light neighbours (O), in the second one the same photoabsorber is surrounded by a larger number (six) of heavy neighbours (I). As stated before, an unique shell around the photoabsorber is considered in both models.

The GNXAS package (Filipponi, DiCicco & Natoli 1995, Filipponi & Di Cicco 1995) has been used to calculate the reference EXAFS of both models. The reference XAS spectrum was built using a Debye-Waller factor equal to $\exp(-2\cdot a^2\cdot k^2)$, with $a^2$ = 0.01Å$^2$ and adding: a) a pre-edge background modelled with a straight line, b) an edge step modelled with an arctg function broadened for the finite core-hole lifetime corresponding to a FWHM of 8.123 eV and with an edge jump set to $J$ = 1 and centred at $E_0$=25516.5 eV; c) a post edge background modelled using the linearized hydrogenic model for the atomic background absorption: $J\cdot(1-8v/3)$ with $v = (E-E_0)/E_0$. The presence of XANES structures on and near the edge was deliberately ignored for the sake of simplicity. Each reference XAS spectrum so obtained will be referred to in the following as $mu_{ref}(E)$.

## 2.2. Synthetic 'experimental' spectra

Starting from each reference spectrum, a set of 50 noisy spectra were simulated using a random number generator for adding to each point of the reference spectrum a Gaussian noise in the range $mu_{ref}(E)(1+/-10^{-4})$ as an approximation of a Poissonian noise for a counting rate of $10^8$ counts/s. We assume that this is a realistic noise level for a "good" experimental spectrum that can be routinely obtained at a third generation synchrotron radiation source.

## 2.3. Fitting procedure

Each synthetic XAS spectrum has been fitted using again the GNXAS package. In particular, the program performs the following preliminary operations (not repeated successively):

1) linear fit of the pre-edge region;
2) calculation of the first derivative of the absorption signal and determination of the edge position from the maximum;
3) preliminary post edge background extraction, fitting the post edge signal with *beta*-splines and without considering the structural parameters.

The program then compares the post edge absorption with a model which is composed by a smooth post-edge background part (*beta*-splines) plus a structural oscillating part which is actually the EXAFS.

## 2.4. Evaluation of structural parameters

A total number of twenty four different sets of fit parameters have been produced in correspondence to each different combination concerning:

1. Kind of model ($AgO_2$ or $AgI_6$),
2. Freely adjustable parameters: a) *n* (coordination number), $a^2$ (Gaussian part of the Debye-Waller factor), *r* (shell radius), $E_{0,T}$ (*k*-space origin), with *J* (edge jump) fixed at its theoretical value of 1, or b) the previous set of adjustable parameters *and J.*
3. Sampling range in *k* space: a) 2 – 12 Å$^{-1}$, or b) 3.5 – 12 Å$^{-1}$, or c) 5 – 12 Å$^{-1}$, for the $AgO_2$ model; a) 2 – 16 Å$^{-1}$, or b) 3.5 – 16 Å$^{-1}$, or c) 5 – 16 Å$^{-1}$, for the $AgI_6$ model.
4. Weighting exponent in *k* space: a) fit of chi (*k*) [which should be the correct choice in the case of a constant noise], or b) fit of $k^3$.chi(*k*) [which corresponds to what is more usually made in the EXAFS community]; *i.e.* fit of $k^m$. chi (*k*) with $m = 0$ (case a) or $m = 3$ (case b).

In the followings, a fit according for a particular case among all these options will be concisely referenced, for instance, with "*J*/2-12/3" to indicate that the pertinent model is fitted including the edge jump among the freely adjustable parameters, sampling the *k* space between 2 and 12 Å$^{-1}$, and using $m = 3$ weights, or with "*J*=1/5-16/0" to indicate that the pertinent model is fitted without including the edge jump among the freely adjustable parameters, sampling the *k* space between 5 and 16 Å$^{-1}$, and using $m = 0$ weights.

## 3. Results and discussion

We start our discussion with the $AgI_6$ model. For this model, the upper part of Fig. 1 gives an example of a single simulated EXAFS (dots) and its fit (continuous line) obtained with the *J*/2-16/3 procedure. The corresponding Fourier transforms are shown in the lower part of the same figure.

By considering the whole set of 50 simulated EXAFS and the same procedure, one obtains a population of 50 $\{J, r, n, E_{0T}, a^2\}$ or $\{r, n, E_{0T}, a^2\}$ parameters. From this population, several statistical estimators (mean values, standard deviations, …) are obtained.

Table 1 summarises the results concerning mean values and standard deviations for the *J*/2-16/3 fit procedure. It might be interesting to note that all parameters (with the notable exception of $E_{0,T}$) are biased by amounts that are much larger than the sample standard deviations. The bias, however, is around 6% for the coordination number, well below 0.01 Å for the coordination distance, and below 2% for the Debye-Waller factor. All these biases are nicely smaller than what is usually believed to be a typical EXAFS error on these parameters. Moreover, the bias is around 10% for the edge jump.

Deeper insights can be obtained by exploring the scatter plots (Fig. 2-6) where the values of couples of parameters obtained from the fit of a particular reference model and according to a fixed fit strategy (again: *J*/2-16/3) are plotted against each other for the whole set of simulated experiments. These figures show, for instance, that edge jump and Debye-Waller factor appear as only weakly correlated variables (fig. 2), while edge jump and coordination number are much more heavily correlated to each other (see Fig. 3). Concerning the latter couple, it is also interesting to note that the fit results are heavily displaced on the scatter plot from the ‘true’ values and that their joint correlation line, when extrapolated to $J = 1$, gives a much more accurate $n$ value. This suggests that taking the edge jump as a fixed parameter produces a much better value for the coordination number.

Fig. 7- 11 compare mean values and standard errors for different fit procedures and for the simulated experiments on $AgI_6$. On these plots, the error bars are the standard errors calculated over the sample of computer simulated experiments, and the horizontal lines are the true values. The $x$ axis is simply a label of the fit procedure and the correspondence is as follows: 1: $J$=1/2-16/3; 2: $J$=1/3.5-16/3; 3: $J$=1/5-16/3; 4: $J$=1/2-16/0; 5: $J$=1/3.5-16/0; 6: $J$=1/5-16/0; 7: $J$/2-16/3; 8: $J$/3.5-16/3; 9: $J$/5-16/3; 10: $J$/2-16/0 11: $J$/3.5-16/0; 12: $J$/5-16/0.

In some detail, Figs. 7 and 8 show the results obtained for $J$ and $n$. When $J$ is fixed to its true value ($x$-labels 1 to 6), the coordination numbers obtained are quite close to the real value, with a bias that is close to the statistical error and well below 20 % (which is commonly believed to be a good estimate of the error in the coordination numbers as determined by EXAFS). In addition, the statistical error is almost independent on the weighting scheme and on the range of $k$ space used in the fit. On the contrary, allowing $J$ to float (cases 7 - 12) introduces a bias in the determination of coordination numbers which is much greater than the statistical spread (but in any case below 20 %).

The correlation between $n$ and $J$ is well known and is due to the fact that they both contribute to the EXAFS as normalising factors. In particular, it is well apparent from Figs. 7 and 8 that, when $J$ is allowed to float, the $n$ values obtained from the fit are almost always greater than the true value, while the $J$ values are always lower. Moreover, $J\cdot n$ (which is the actual normalizing factor of the EXAFS) is regularly underestimated (from the mean values of Table 1, $\langle J\cdot n\rangle = 5.85$ instead of 6). In our opinion, this last fact is due to broadening of the edge caused by the finite lifetime of the core hole. The broadened edge jump is then accounted by the *beta*-splines used for fitting the post-edge background in form of a lower trend of the average signal just after the edge. Then, the extrapolation to the edge energy of this ‘wrong’ trend produces an underestimation of $J$ (and $J{\cdot}n$) which is calculated as the difference between post-edge and pre-edge backgrounds at the edge. An indirect proof of the above understanding has been obtained by considering a restricted number of simulated experiment based on a reference EXAFS with a stepped edge jump (instead of the rounded arctg shape). In this case, the fit gives much more accurate parameters: $<J> = 0.98$, $<n> = 6.07$ and $<J{\cdot}n> = 5.95$.

The $a^2$ parameter (see Fig. 9) is almost independent on the actual fit conditions and is always quite well determined, both the bias and the statistical error being of the order of few percents. This despite the fact that there are correlations of $a^2$ both with $J$ and $n$. Seemingly, what is important here is the *total* correlation of all the three parameters $a^2$, $J$ and $n$; $J$ and $n$ may well be heavily biased, but their combination is accurate enough to allow an accurate determination of $a^2$.

The $E_{0T}$ parameter is seemingly well recovered (Fig. 10) for all fit procedures. Concerning the $r$ parameter (Fig. 11), the bias is well above the statistical error. $E_{0T}$ and $r$ are strongly correlated, but this correlation is not enough to explain the bias on $r$. We think that this is (at least partially) an indirect effect of the disorder ($a^2$), because in calculating the mean geometrical configuration the fitting program uses other factors in addition to the simple Debye-Waller weight coming from the $a^2$ factor: this produces small phase and amplitude shifts which possibly are at the origin of the bias experimentally observed. Of course, this second order effect is expected to vanish in the limit $a^2 \to 0$. Indeed, using a reference model with a ten times lower $a^2$ we actually obtained a six times lower bias on *r*. It should be noted, however, that the (systematic) error affecting *r* is of the order of a few thousandths of Å, that is well below the value (1-2 hundredths of Å) which is usually assumed as a reasonable estimate of the error in the determination of the bond distances by EXAFS.

Finally, for what concerns more generally the correlations between different parameters, it can be said that the case of the *J*/2-16/3 procedure is well representative of the results for the different procedures.

The results for the $AgO_2$ model are shown in Fig.s 12-16. The results for this model are *qualitatively* similar to those of the $AgI_6$ model, but in general correlations, biases and the statistical errors are greater. For example, the systematic error which affects *r* can be of the order of 1-2 hundredths of Å, and that which affects *n* can reach the value of 20 % if *J* is allowed to float. This result is expected and is due to the much lower level of the EXAFS oscillations for this model.

## 4. Conclusions

The most important result of the present simulation approach is that random noise plays a minor role with respect to systematic errors in determining accuracy, precision and correlation of the structural parameters obtained from analysis of an EXAFS spectrum. This inference is well apparent when one considers, on one side, that the simulated experiments only account for random noise and, on the other side, that for most parameters the statistical spread here obtained is typically much lower than what is reported in the scientific literature as the typical error of a real EXAFS analysis. The conclusion is also supported by the simulation results alone, by considering the significant difference (for some parameters) between bias and statistical spread.

If we now remind that the background is here modelled only in a drastically simplified way, while a real EXAFS spectrum is much more difficult to analyse on this regard, it seems reasonable to infer that the most important source of error in the analysis of an EXAFS spectrum is related to the procedure of (pre-edge and/or post-edge) background subtraction. On this regard, the computer simulation also makes clear the importance of the *a priori* knowledge of the correct value of the edge jump ($J$), which is essential for an accurate determination of $n$. It is therefore suggested to start an EXAFS analysis only after having obtained an independent and reasonably accurate value of the edge jump, for instance by comparison with the spectra of proper standards of precisely known coordination number or, more directly, by accurately determining the amount of photo-absorber atoms in the sample.

Finally, the present simulation approach show that the accuracy of the fitted parameters is practically independent both of the weighting scheme and of the fitted range in $k$ space. The latter aspect practically affects only the determination of a correct value of $J$ if – contrary to the suggested procedure – one wants to keep this parameter free to change during fitting.

## Acknowledgements

The Authors want to thank Adriano Filipponi (Università dell’Aquila, Italy) for his deep insight into the details of the GNXAS programme and for stimulating discussions on aim and results of the present work.

This work has been partially supported by the Department of University and Scientific and Technological Research of the Italian Government (MURST-40%).

# References

Binsted, N., Gurman, S. J., Campbell, T. C, and Stephenson, P. C.(1998): EXCURV98: SERC - Daresbury Laboratory.

Curis, E. and Benazeth, S. (2000). *J. Synchrotron Rad.* **7**, 262.

Dapiaggi, M., Anselmi-Tamburini, U., and Spinolo, G. (1998). *J.Appl.Cryst.* **31,**379-387.

Filipponi, A., Di Cicco, A., and Natoli, C.R. (1995). *Phys.Rev.* **B 52,**15122.

Filipponi, A. and Di Cicco, A. (1995). *Phys.Rev.* **B52,**15135.

Filipponi, A. (1995). *J.Phys.Condensed Matter* **7,**9343.

Incoccia, L. and Mobilio, S. (1984). *Il Nuovo Cimento,* **3 D** (5), 867.

James, F. & Roos, M., CERN Computer Centre Program Library, Program D 506

Krappe, H. J. and Rossner, H. H. (2000). *Phys Rev.* **B 61**, 6596.

Lytle, F.W., Sayers, D.E. , and Stern, E.A. (1989). *Physica* **B 158,** 701.

Press W. H., Flannery B. P., Teukolsky S. A. & Vetterling W. T. (1988). Numerical recipes in C, Cambridge University Press.

Rehr, J.J., Booth, C.H., Bridges, F., and Zabinsky, S.I. (1994). *Phys.Rev.* **B49,**12347.

Rehr, J. J. (2000). *Rev. Mod. Phys*. **72**, 621.

## Table 1

Mean values and standard errors over the sample of simulated experiments obtained on the $AgI_6$ model with the *J*/2-16/3 fit procedure.

| Parameter | True value | Mean value from simulation | Std. error |
|---|---|---|---|
| $n$ | 6. | 6.43 | 0.02 |
| $r$ (Å) | 3.1 | 3.1059 | $2 \cdot 10^{-4}$ |
| $E_{0,\mathrm{T}}$ (eV) | 25516.5 | 25516.49 | 0.01 |
| $J$ | 1 | 0.9082 | $0.8 \cdot 10^{-4}$ |
| $a^2$(Å$^2$) | 0.01 | $9.86 \cdot 10^{-3}$ | $4 \cdot 10^{-5}$ |

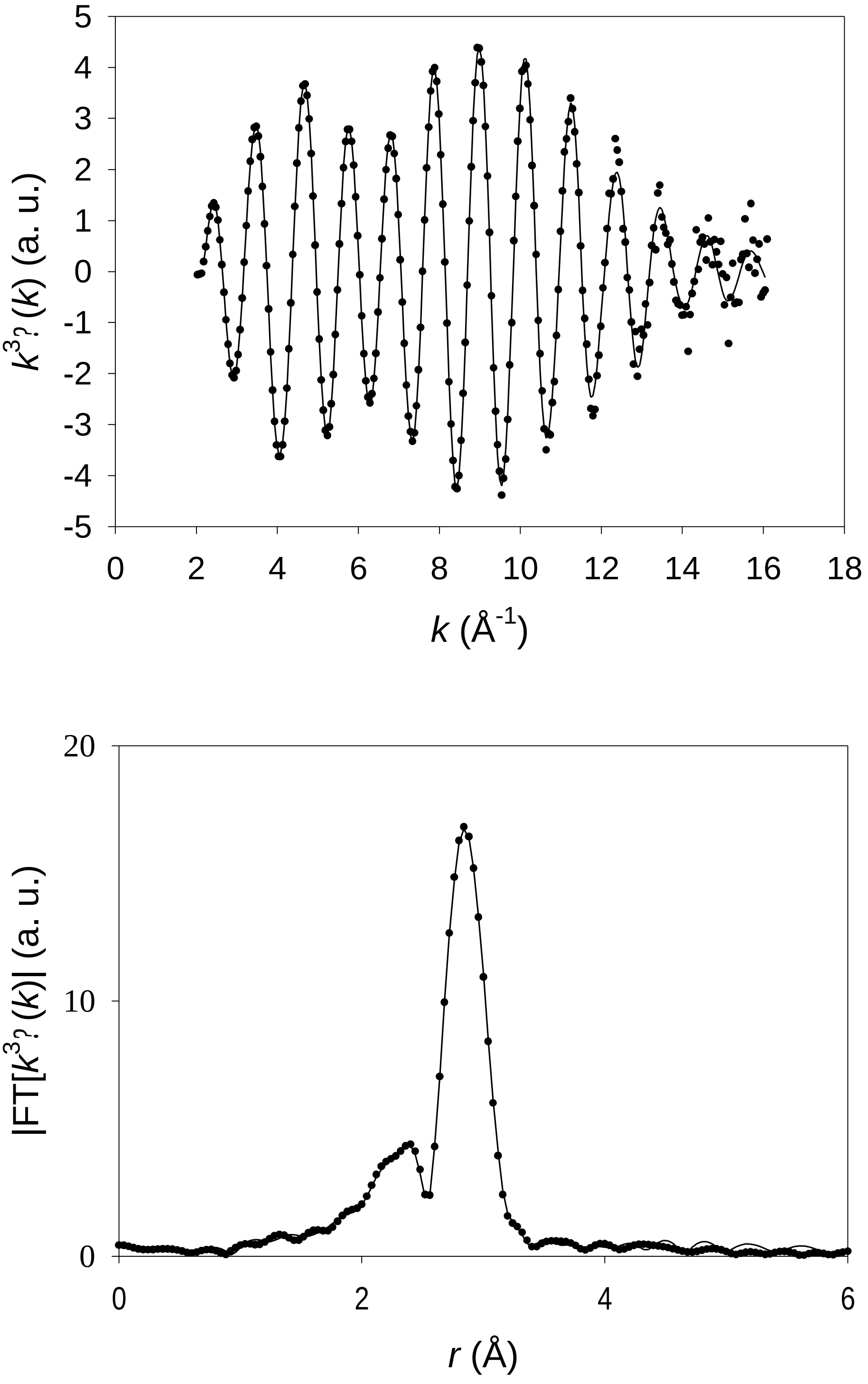

Fig. 1- Example of a single simulated EXAFS (dots) and its fit (continuos line) obtained as described in the text (upper panel). The lower panel shows the modulus of the corresponding Fourier Transform.

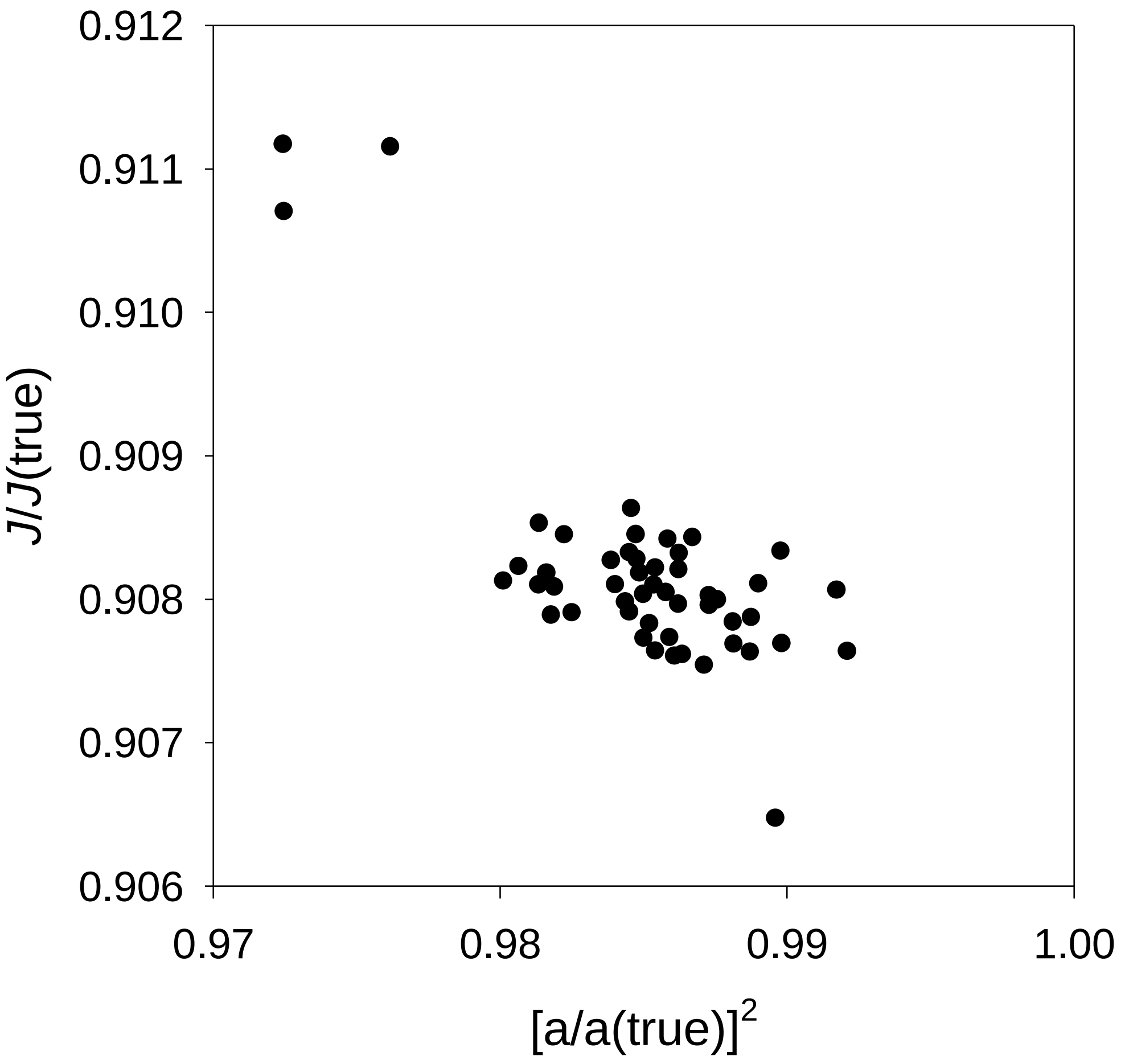

Fig. 2- Correlation between the single pattern evaluations of two parameter values [Gaussian part of the Debye-Waller factor ($a^2$) and edge jump ($J$)]. Each parameter is normalized using its true value.

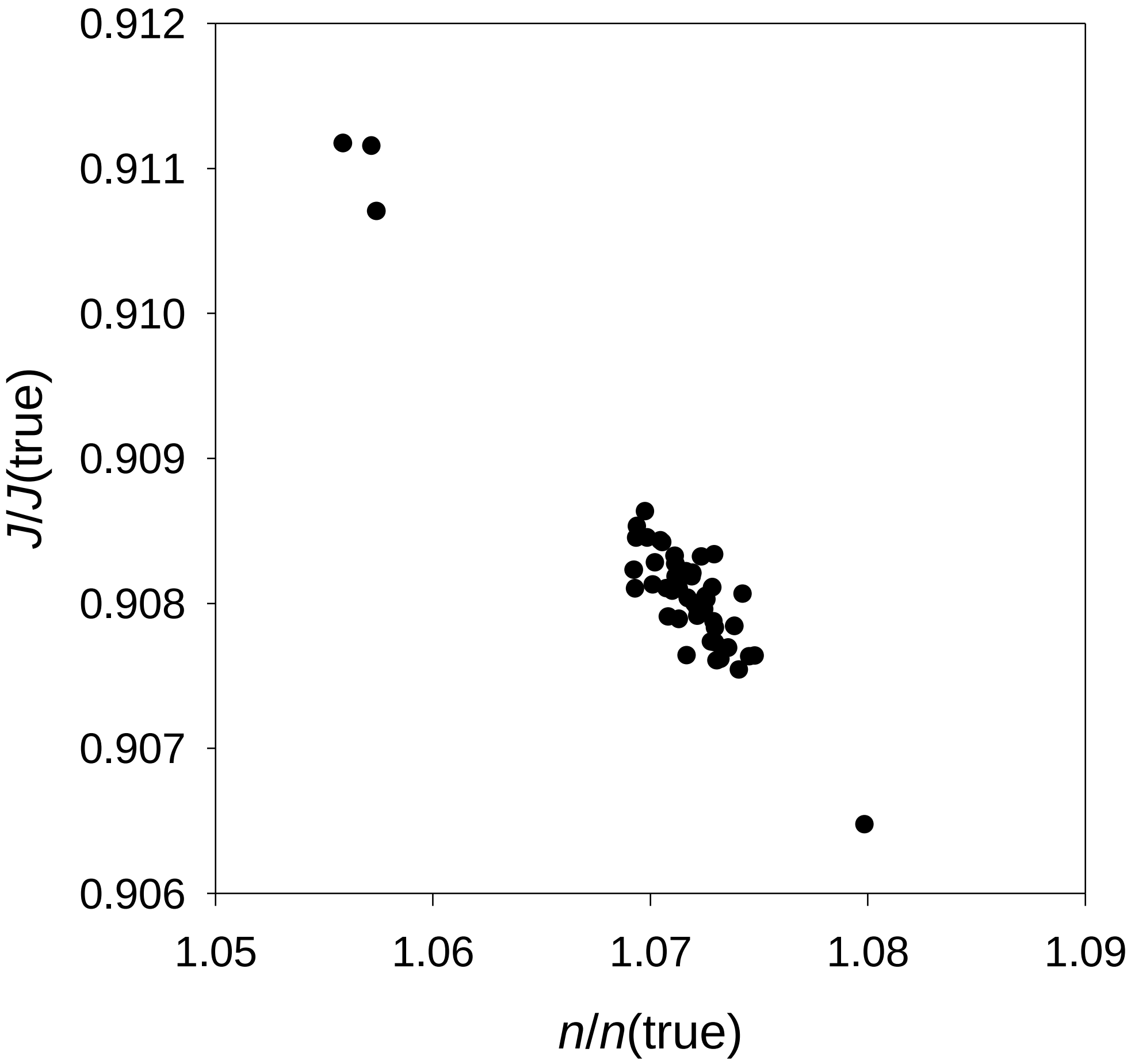

Fig. 3- Correlation between the single pattern evaluations of two parameter values [coordination number (*n*) and edge jump (*J*)]. Each parameter is normalized using its true value.

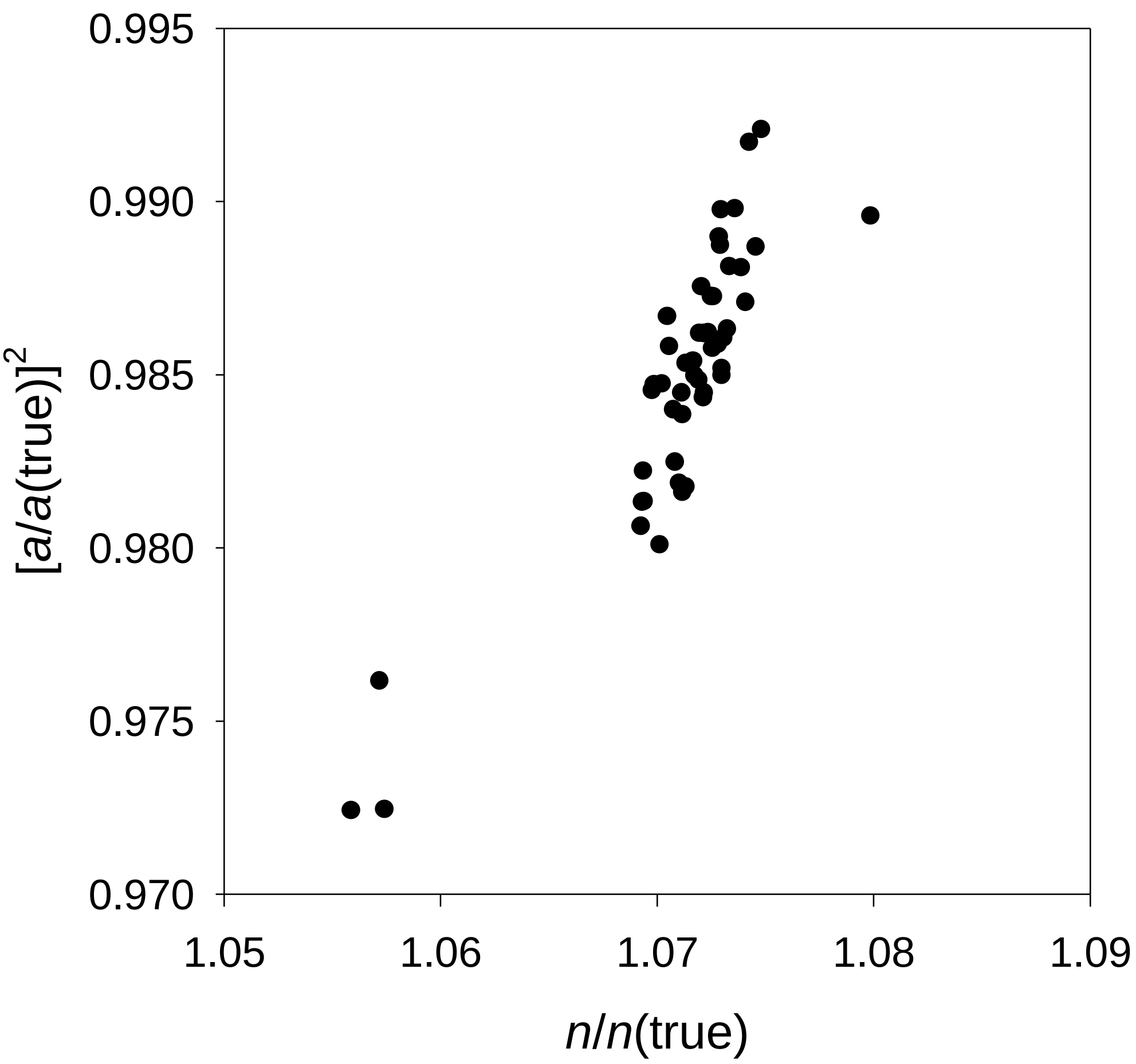

Fig. 4 - Correlation between the single pattern evaluations of two parameter values [coordination number (*n*) and Gaussian part of the Debye-Waller factor ($a^2$)]. Each parameter is normalized using its true value.

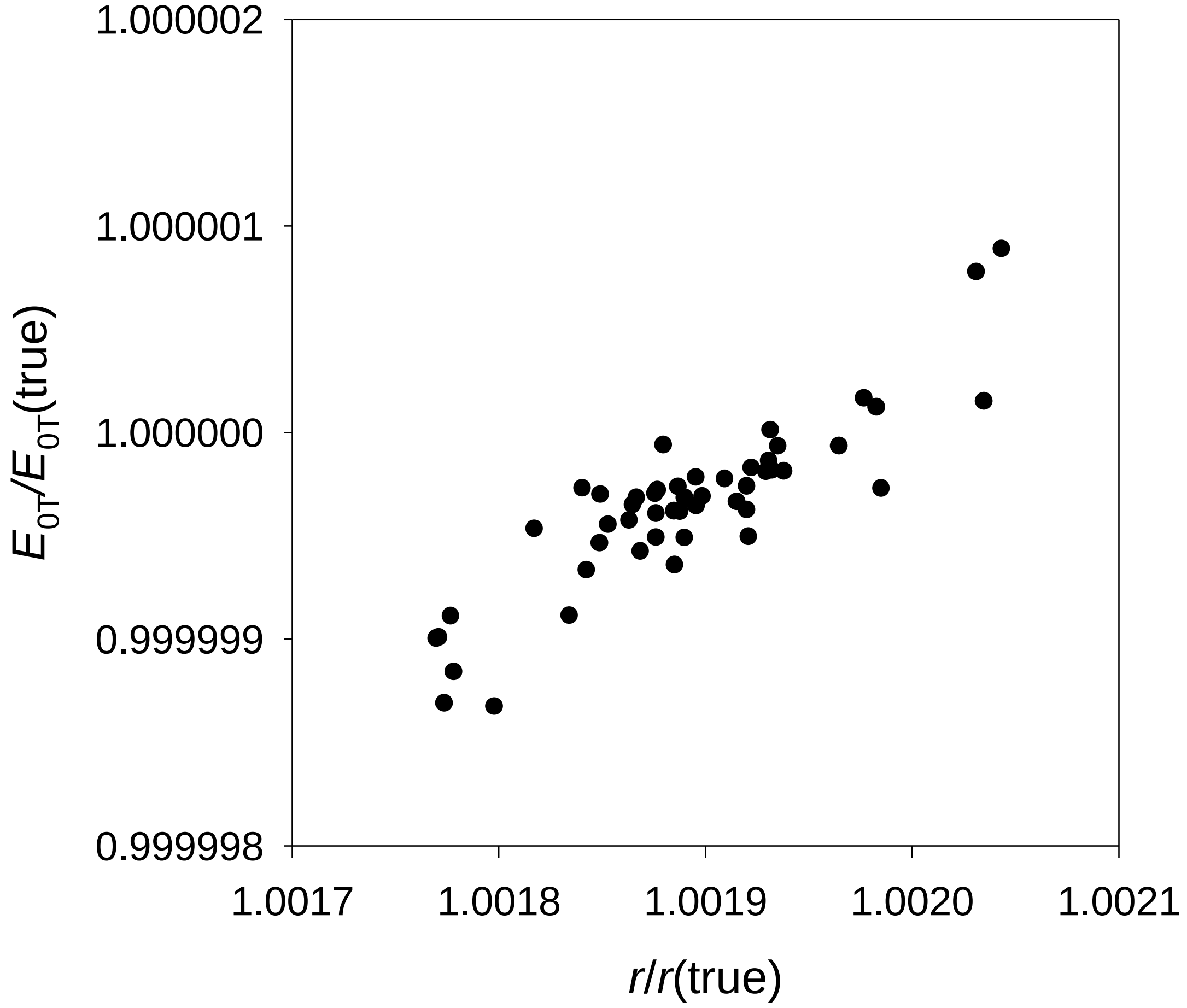

Fig. 5 - Correlation between the single pattern evaluations of two parameter values [shell radius ($r$) and origin of the calculated EXAFS ($E_{0T}$)]. Each parameter is normalized using its true value.

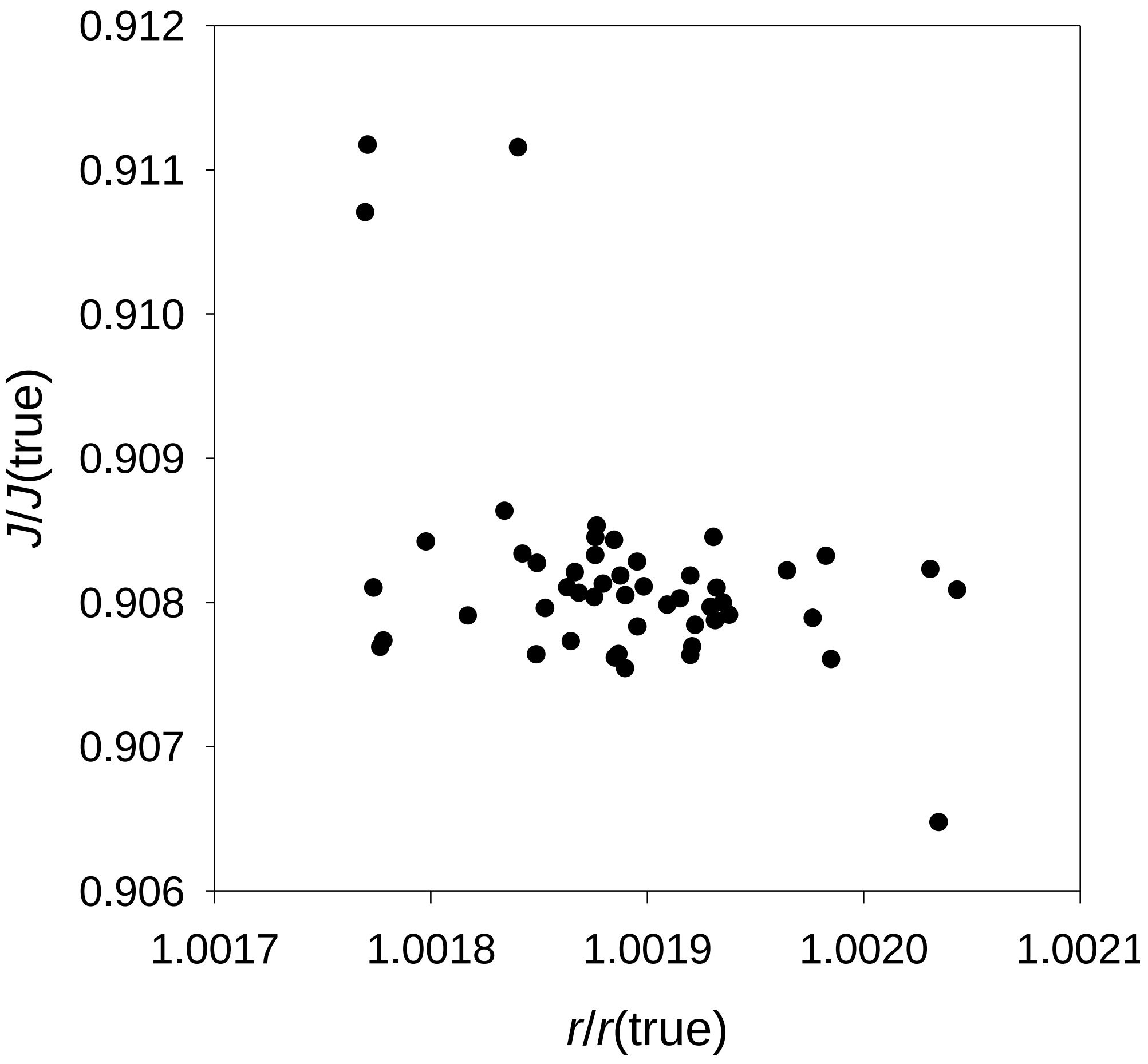

Fig. 6 - Correlation between the single pattern evaluations of two parameter values [shell radius (*r*) and edge jump (*J*)]. Each parameter is normalized using its true value.

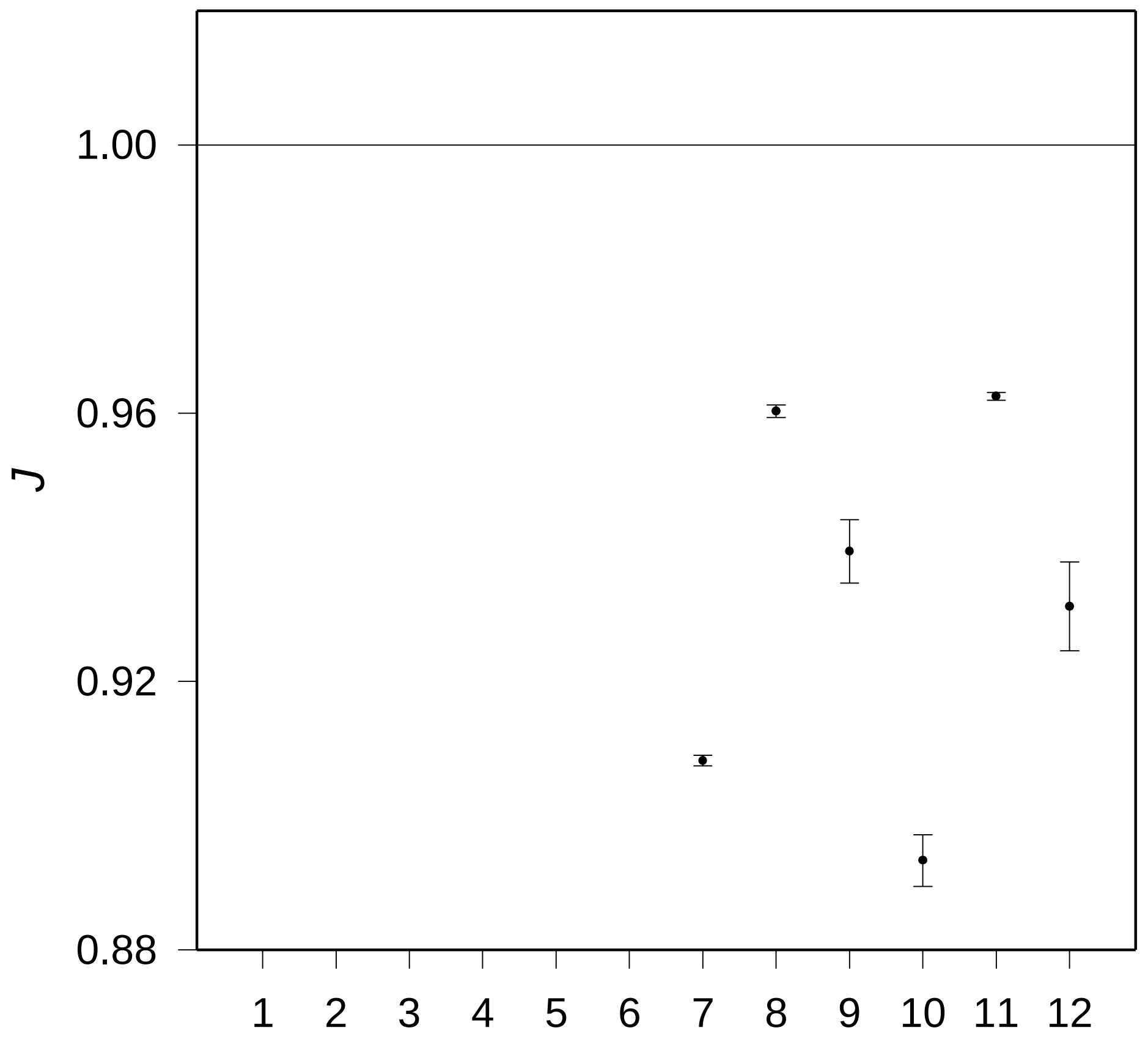

Fig. 7- Edge jump *J* as determined from the different fitting procedures (*x*-labels, see text), for the $AgI_6$ model. The horizontal line marks the true value, and the error bars are the statistical errors calculated over the sample of 50 synthetic experiments. The *J* parameter has been kept fixed at its true value = 1 in the first six fitting procedures.

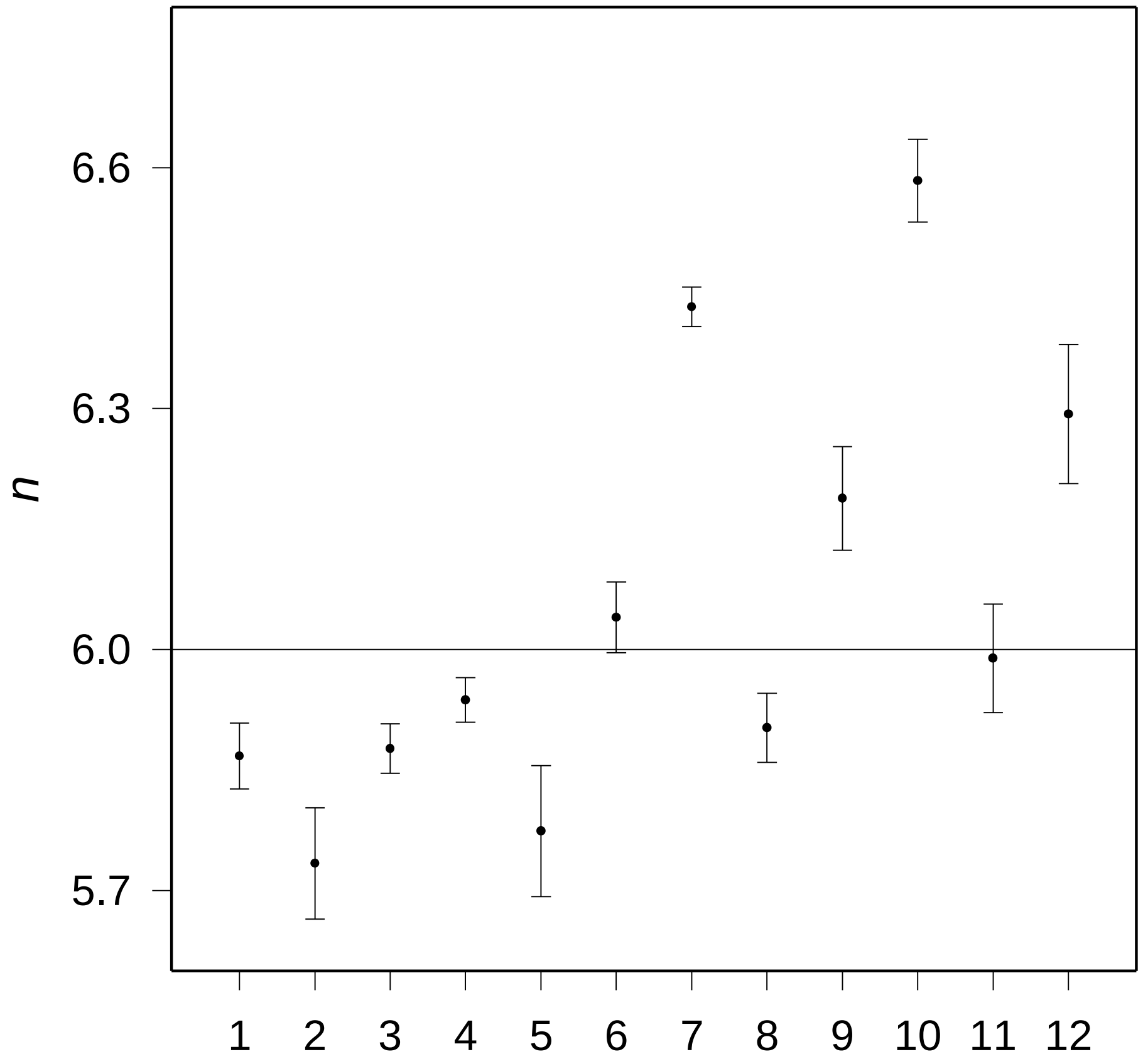

Fig. 8- Same plot as in Fig. 7, but for the coordination number *n*.

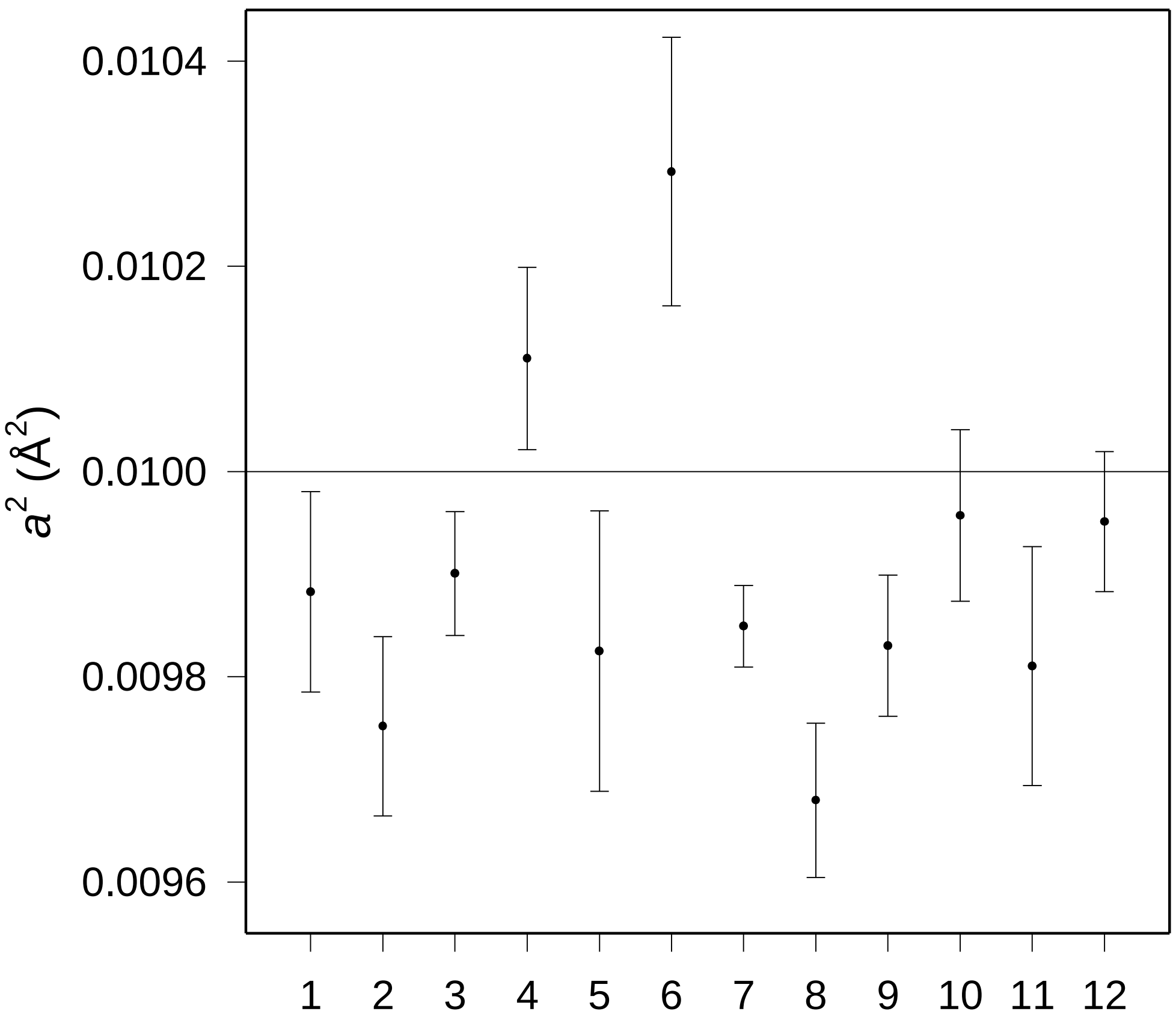

Fig. 9- Same plot as in Fig. 7, but for the Gaussian part of the Debye-Waller factor $a^2$.

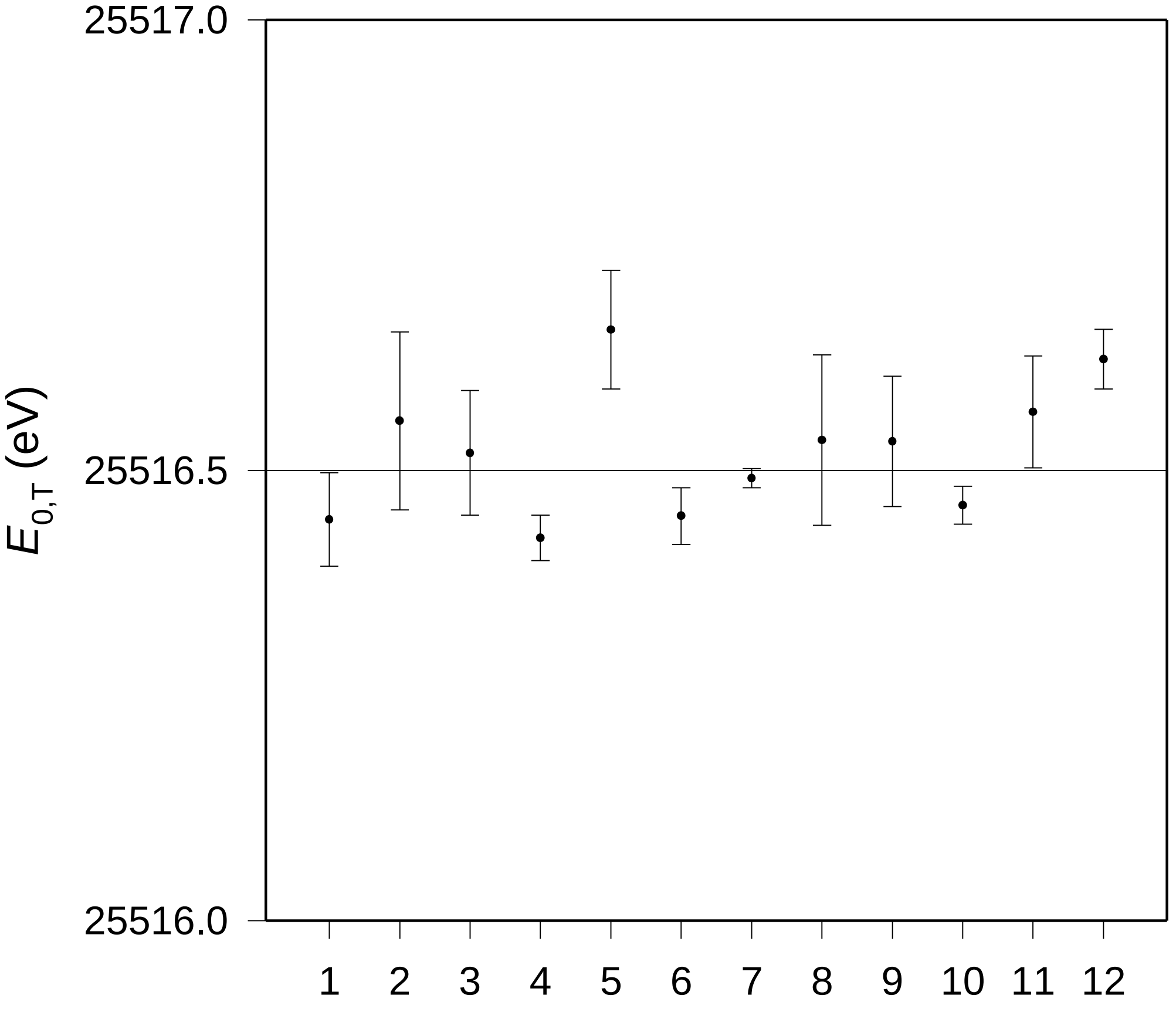

Fig. 10- Same plot as in Fig. 7, but for the origin of the calculated EXAFS ($E_{0T}$).

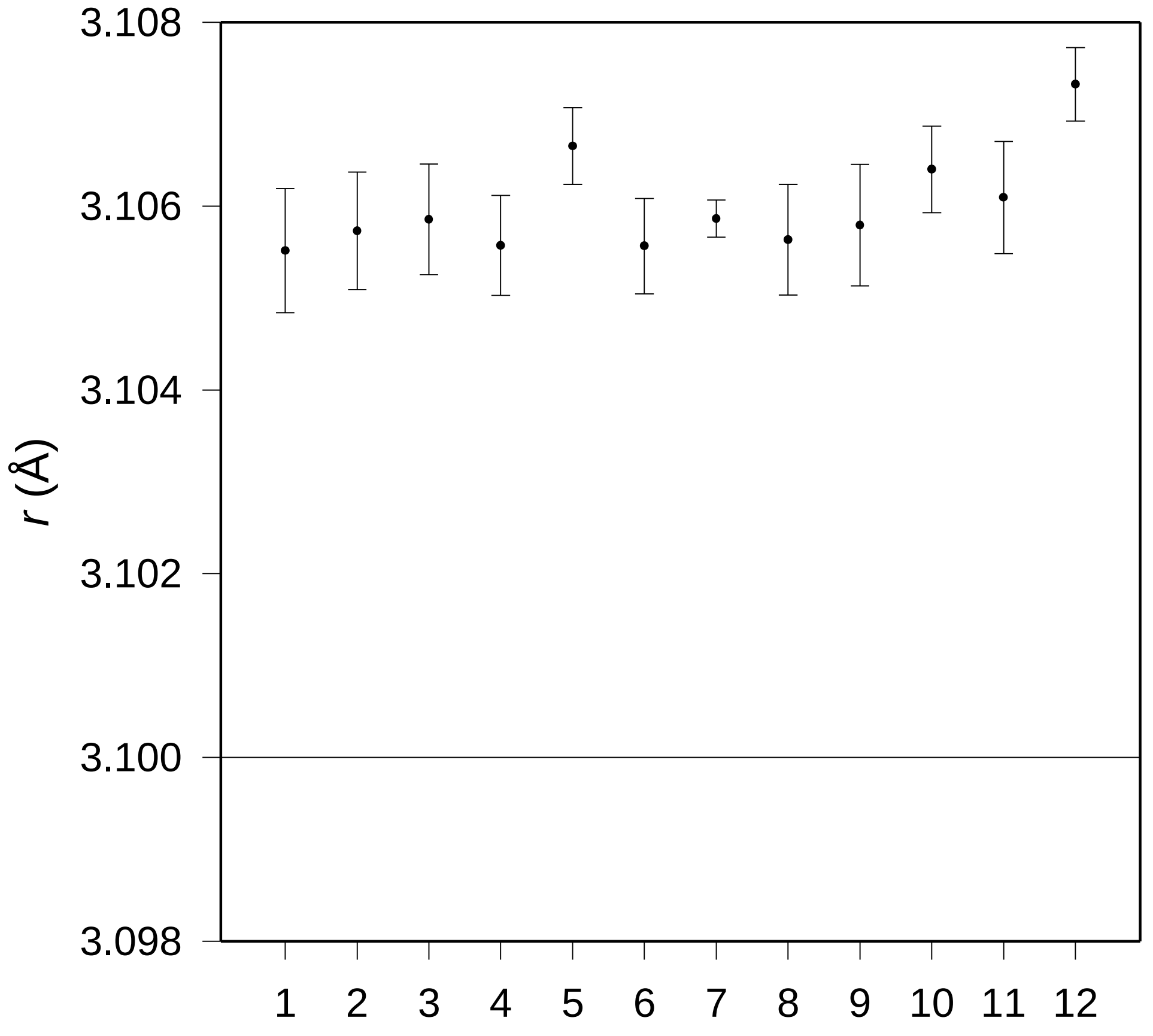

Fig. 11- Same plot as in Fig. 7, but for the shell radius *r*.

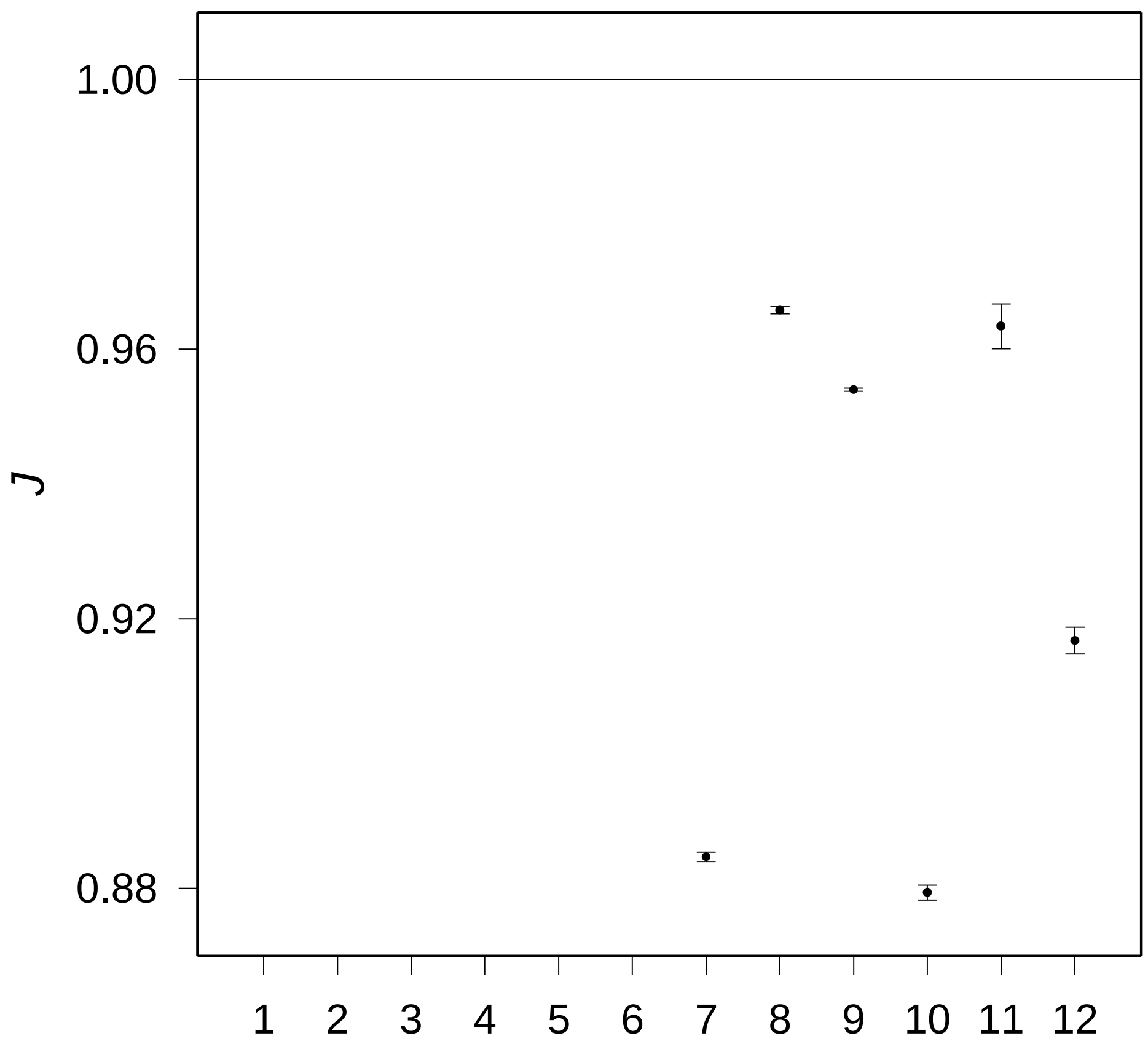

Fig. 12- Same plot as in Fig. 7, but for the $AgO_2$ model.

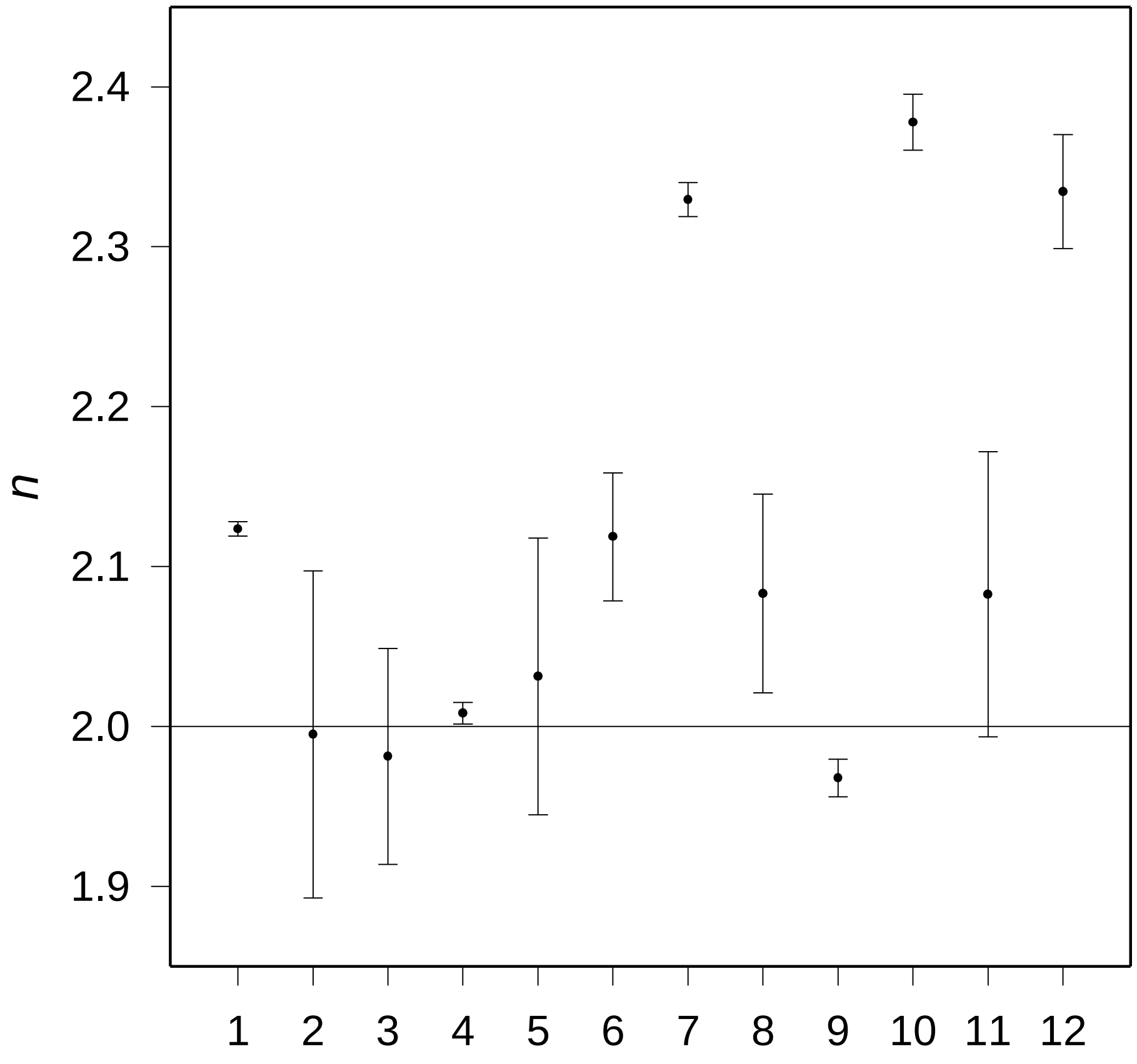

Fig. 13- Same plot as in Fig.8, but for the $AgO_2$ model.

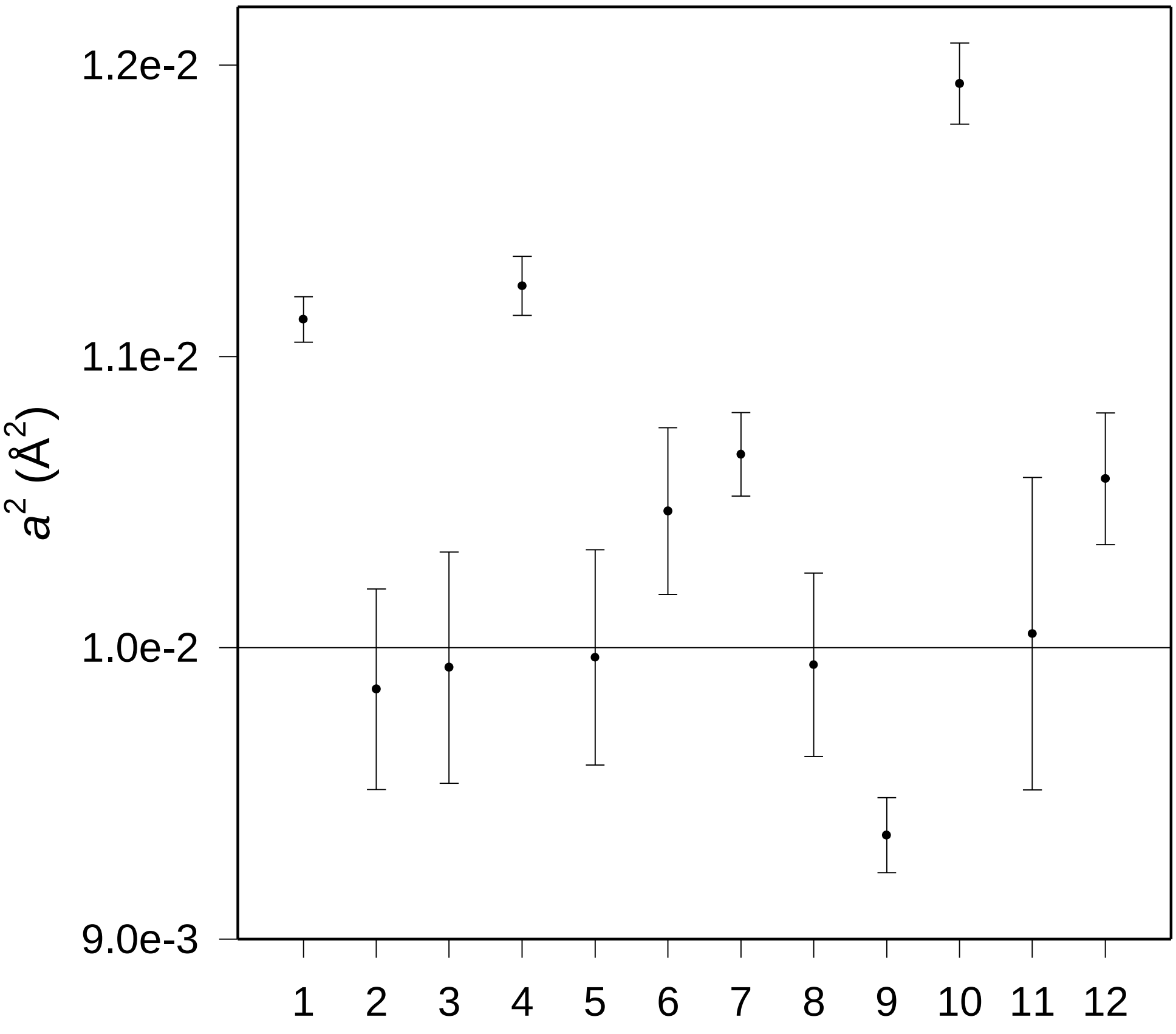

Fig. 14- Same plot as in Fig. 9, but for the $AgO_2$ model.

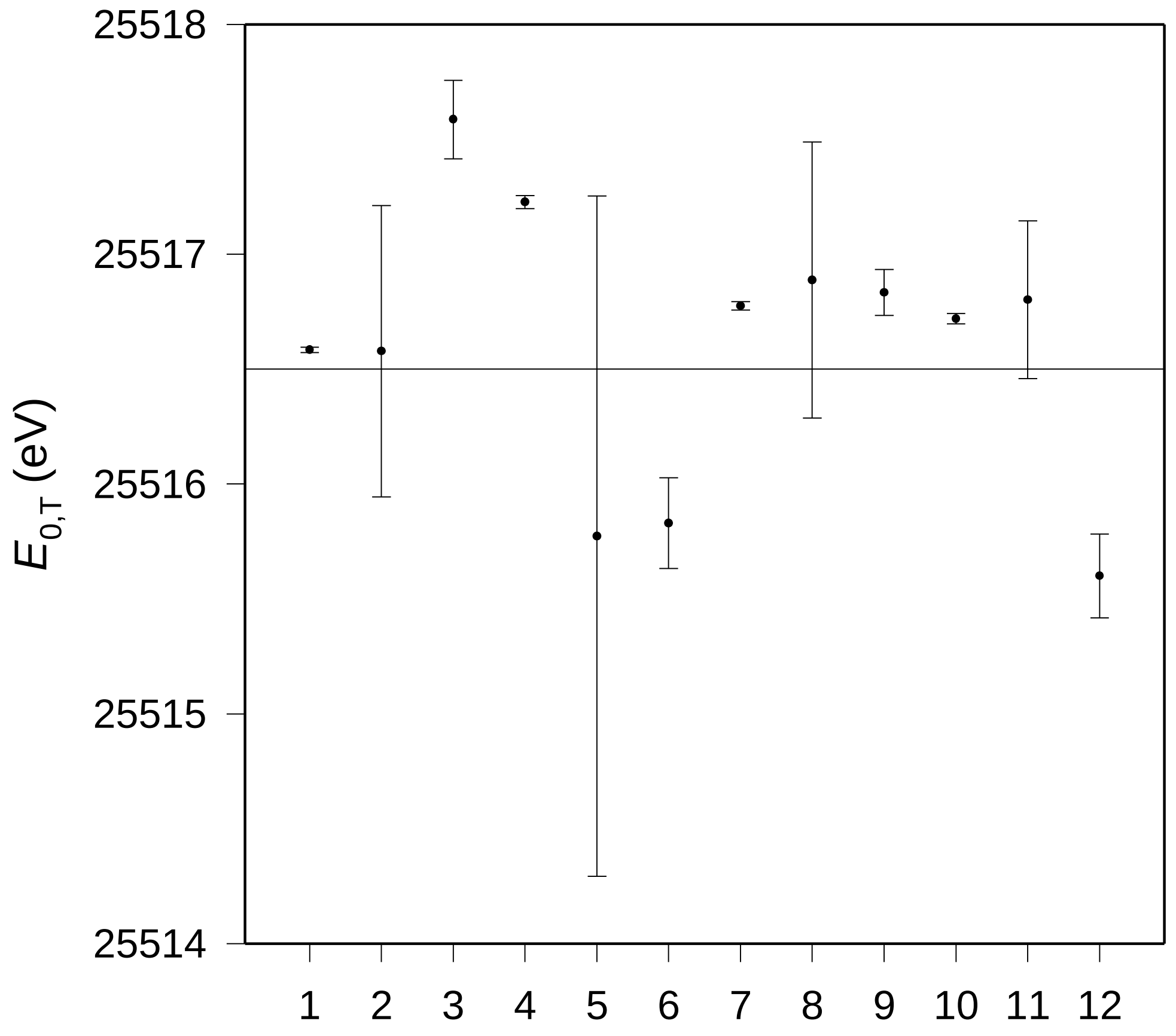

Fig. 15- Same plot as in Fig. 10, but for the $AgO_2$ model.

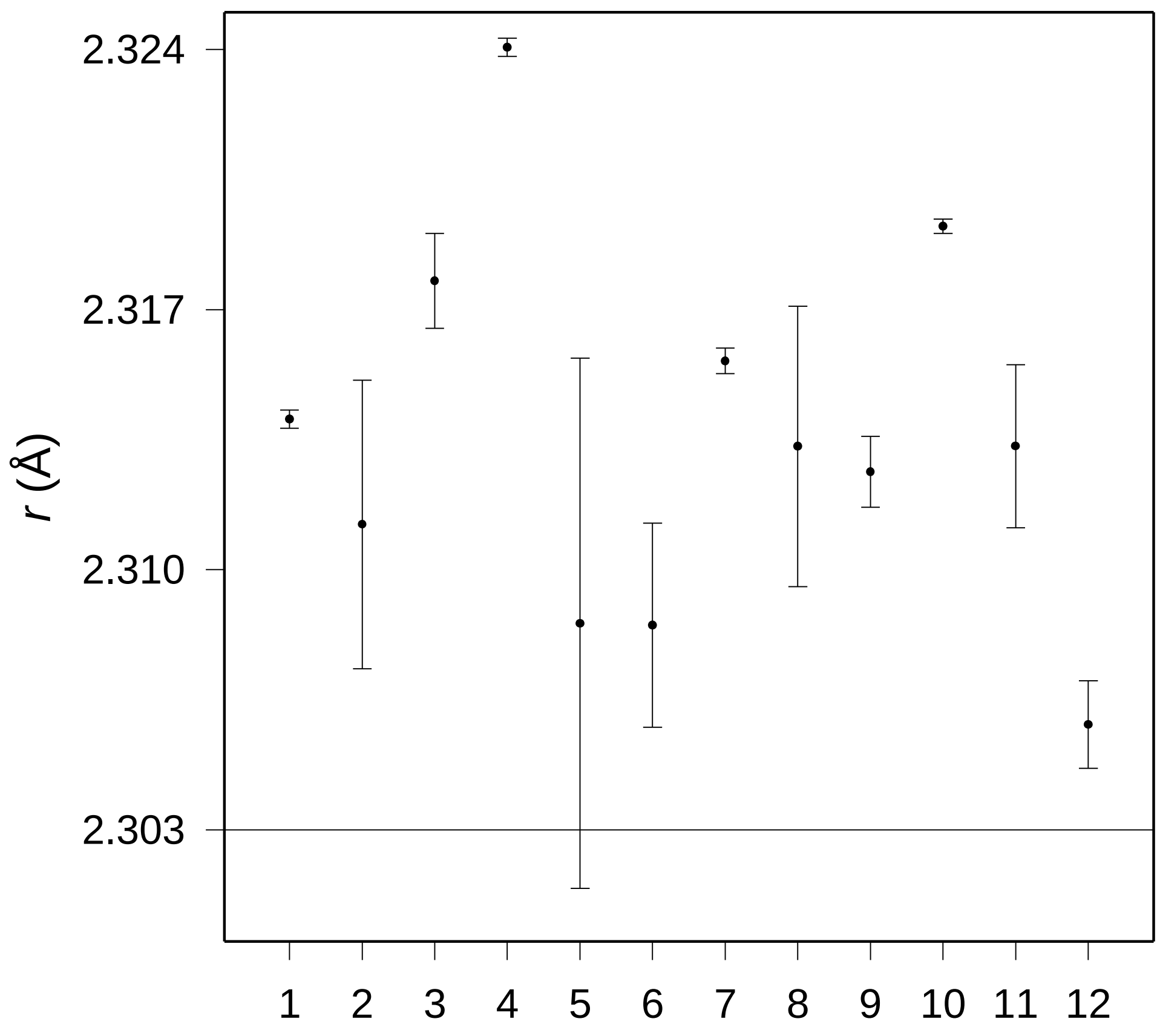

Fig. 16- Same plot as in Fig. 11, but for the $AgO_2$ model.